\documentclass[twocolumn, prX, amssymb]{revtex4-1}
\usepackage{graphicx}% Include figure files
\usepackage{amssymb}
\usepackage{bm}% bold math

\begin{document}

\newcommand{\kbar}{$\bar{K}$}
\newcommand{\km}{$K^-$}
\newcommand{\mMeV}{MeV/$c^2$}
\newcommand{\mGeV}{GeV/$c^2$}

\title{Indication of a deeply bound compact $K^-pp$ state\\
formed in the $pp \rightarrow p \Lambda K^+$ reaction at 2.85 GeV}

 \author{
 T.~Yamazaki$^{1,2}$, M.~Maggiora$^3$, P.~Kienle$^{4,5}$, K.~Suzuki$^{4}$,  
A.~Amoroso$^3$, 
M.~Alexeev$^3$, 
F.~Balestra$^3$,
Y.~Bedfer$^{6}$, %\footnote{Present address: DAPNIA/SPhN, CEA Saclay, F}, 
R.~Bertini$^{3,6}$,
L.~C.~Bland$^{7}$, %\footnote{Present address: BNL, U.S.A.},
A.~Brenschede$^{8}$, %\footnote{Present address: DIAMOS AG, Sulzbach, D}, 
F.~Brochard$^{6}$,
M.~P.~Bussa$^3$, 
Seonho~Choi$^{7}$, %\footnote{Present address: Seoul National University, Seoul, KR},
M.~L.~Colantoni$^3$,
R.~Dressler$^{9}$, %\footnote{Present address: Paul Scherrer Institut, Villigen, CH},
M.~Dzemidzic$^{7}$, %\footnote{Present address: IU School of Medicine, Indianapolis, USA}, 
J.-Cl.~Faivre$^6$,
L.~Ferrero$^3$, 
J.~Foryciarz$^{10,11}$, %\footnote{Present address: Motorola Polska Software Center, Krak\'ow, PL},
I. Fr\"ohlich$^{8}$, %\footnote{Present address: IKF, Frankfurt, D},
V.~Frolov$^{9}$, %\footnote{Present address: Dip. di Fisica Generale and INFN, Torino, I},
R.~Garfagnini$^3$, 
A.~Grasso$^3$,
S.~Heinz$^{3,6}$, %\footnote{Present address: GSI, Darmstadt, D},
W.~W.~Jacobs$^7$, 
W.~K\"uhn$^8$, 
A.~Maggiora$^3$,
D.~Panzieri$^{12}$, 
H.-W.~Pfaff$^{8}$,
G.~Pontecorvo$^{3,9}$, 
A.~Popov$^9$,
J.~Ritman$^{8}$, %\footnote{Present address: FZ, Juelich, D},
P.~Salabura$^{10}$, 
S.~Sosio$^3$,
V.~Tchalyshev$^9$,
and S.~E.~Vigdor$^{7}$, %\footnote{Present address: BNL, U.S.A.} 
}

\address{$^{1}$ Department of Physics, University of Tokyo, Tokyo, 116-0033 Japan}
\address{$^{2}$ RIKEN Nishina Center, Wako, Saitama, 351-0198 Japan}
\address{$^3$ Dipartimento di Fisica Generale ``A. Avogadro'' and INFN, Torino, Italy}
\address{$^{4}$ Stefan Meyer Institute for Subatomic Physics, Austrian Academy of Sciences, Vienna, Austria}
\address{$^{5}$ Excellence Cluster Universe, Technische Universit\"at M\"unchen, Garching, Germany}
\address{$^6$ Laboratoire National Saturne, CEA Saclay, France}
\address{$^7$ Indiana University Cyclotron Facility, Bloomington, Indiana, U.S.A.}
\address{$^8$ II. Physikalisches Institut, Universit\"at Gie\ss{}en, Germany}
\address{$^9$ Forschungszentrum Rossendorf, Germany}
\address{$^{10}$ M.~Smoluchowski Institute of Physics, Jagellonian University, Krak\'{o}w, Poland}
\address{$^{11}$ H. Niewodniczanski Institute of Nuclear Physics, Krak\'ow, Poland}
\address{$^{12}$ Universit\`{a} del Piemonte Orientale and INFN, Torino, Italy}

%%%%%%%%%%%
\thanks{} 
\date{Revised, \today}
% The correct dates will be entered by the editor

\begin{abstract}
We have analyzed data of the DISTO experiment on the exclusive $pp \rightarrow p \Lambda K^+$ reaction at 2.85 GeV to search for a strongly bound compact $K^-pp ~(\equiv X)$ state to be formed in the $pp \rightarrow K^+ + X$ reaction. The observed spectra of the $K^+$ missing-mass and the $p \Lambda$ invariant-mass with high transverse momenta of $p$ and $K^+$ revealed a broad distinct peak with a mass $M_X = 2265 \pm 2~(stat) \pm 5~(syst)$ MeV/$c^2$ and a width $\Gamma_X = 118 \pm 8~(stat) \pm 10~(syst)$ MeV.

%\keywords{$\bar{K}$ nuclei \and kaon condensation \and super-strong nuclear force \and strange dibaryon}
% \PACS{PACS code1 \and PACS code2 \and more}
% \subclass{MSC code1 \and MSC code2 \and more}
\end{abstract}

\pacs{13.75.Jz 21.45.+v  21.85.+d 21.90.+f  24.10.-i  21.30.Fe 25.40.-h }% PACS, the Physics and Astronomy

\maketitle

Recently, it was predicted that a strongly bound $K^-pp$ system with a short $p$-$p$ distance \cite{Akaishi:02,Yamazaki:02} can be formed in a $p+p \rightarrow p + \Lambda^* + K^+$ reaction with an enormously large sticking probability between $\Lambda^* \equiv \Lambda(1405)$ and $p$ due to the short range and high  momentum transfer of the $pp$ reaction \cite{Yamazaki:07a,Yamazaki:07b}. Thus, the issue whether the $K^-pp$ and other kaonic nuclei have ultra high nuclear density or not can be answered by examining this prediction experimentally. Here, we report that existing data of a DISTO experiment show an evidence for this ``exotic" formation. Preliminary reports have been published \cite{Yamazaki:09,Maggiora:09}.

$K^-pp$ (a symbolical notation; more generally, $(\bar{K} NN)_{S=0,I=1/2}$) is the simplest kaonic nuclear bound system, predicted to be a quasi-stable state with a mass $M = 2322$ MeV/$c^2$, a binding energy $B_K = 48$ MeV and a partial decay width $\Gamma_{\Sigma \pi p} = 61$ MeV \cite{Akaishi:02,Yamazaki:02}. A detailed theoretical analysis, based on the Ansatz (called ``strong binding regime") that the $\Lambda (1405)$ resonance is an $I=0$ $\bar{K} N$ quasi-bound state embedded in a $\Sigma \pi$ continuum, has shown that $K^-pp$ has a molecule-like structure in which the $K^-$ migrates between the two protons, causing a {\it super-strong nuclear force} \cite{Yamazaki:07a,Yamazaki:07b,Ivanov:09}. The strongly bound nature of $K^-pp$ is also supported by recent Faddeev calculations \cite{Shevchenko,Ikeda}. On the other hand, recent theories based on chiral dynamics predict the $K^-p$ pole to be located at 1420-1430 MeV/$c^2$, close to the $\bar{K} + N$ threshold, leading to a ``weak" $\bar{K} N$ interaction \cite{Oset,Weise:07} (``weak-binding regime"), and thus to a very shallow $K^-pp$ state \cite{Dote:09}. Since the issue is pertinent to the existence of dense kaonic nuclear states \cite{Akaishi:02,Yamazaki:02,Dote:04,Yamazaki:04} and eventually to the problem of kaon condensation \cite{Kaplan}, it is of vital importance to distinguish between the ``strong binding" and the ``weak binding" regimes by studying the $K^-pp$ formation in the $pp$ reaction experimentally.
 So far, only little experimental information is available. A FINUDA experiment of $K^-$ absorption at rest in light nuclei \cite{Agnello:05} reported that the observed invariant-mass spectrum, $M(p \Lambda)$, showed a peak-like structure with $M = 2255 \pm 7$ MeV/$c^2$, $B_K = 115 \pm 7$ MeV and $\Gamma = 67 \pm 14$ MeV/$c^2$, but unfortunately, its lower part is suppressed by the detector acceptance and no information for the shape of possible physical background is available. 

\begin{figure}[t]
  \begin{center}
     \includegraphics[height=4.1cm]{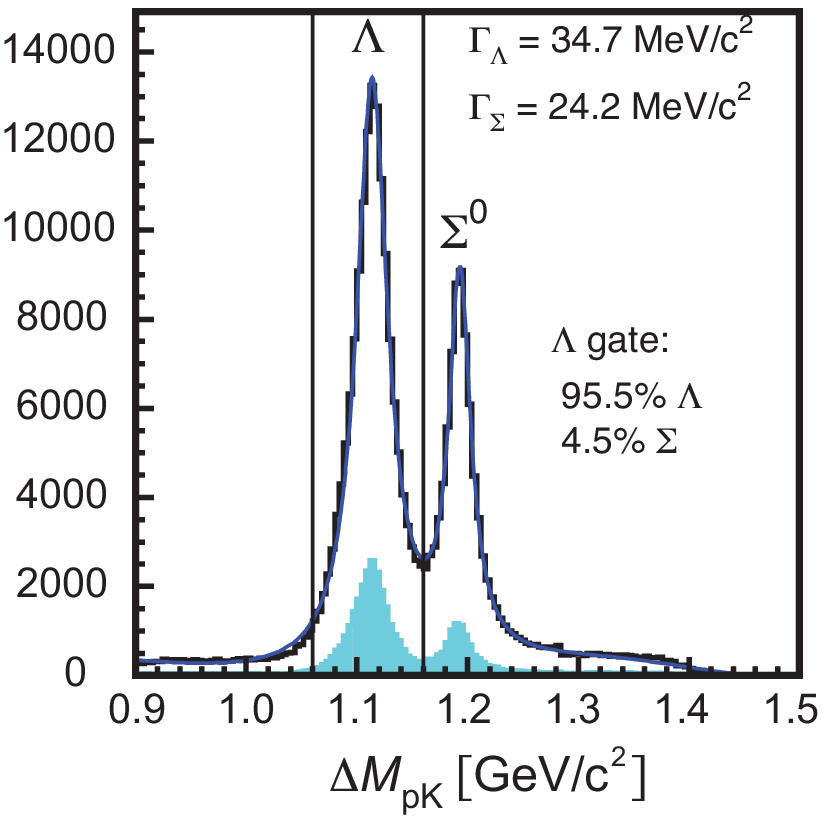}% prc_2010mm_resol
      \includegraphics[height=4.1cm]{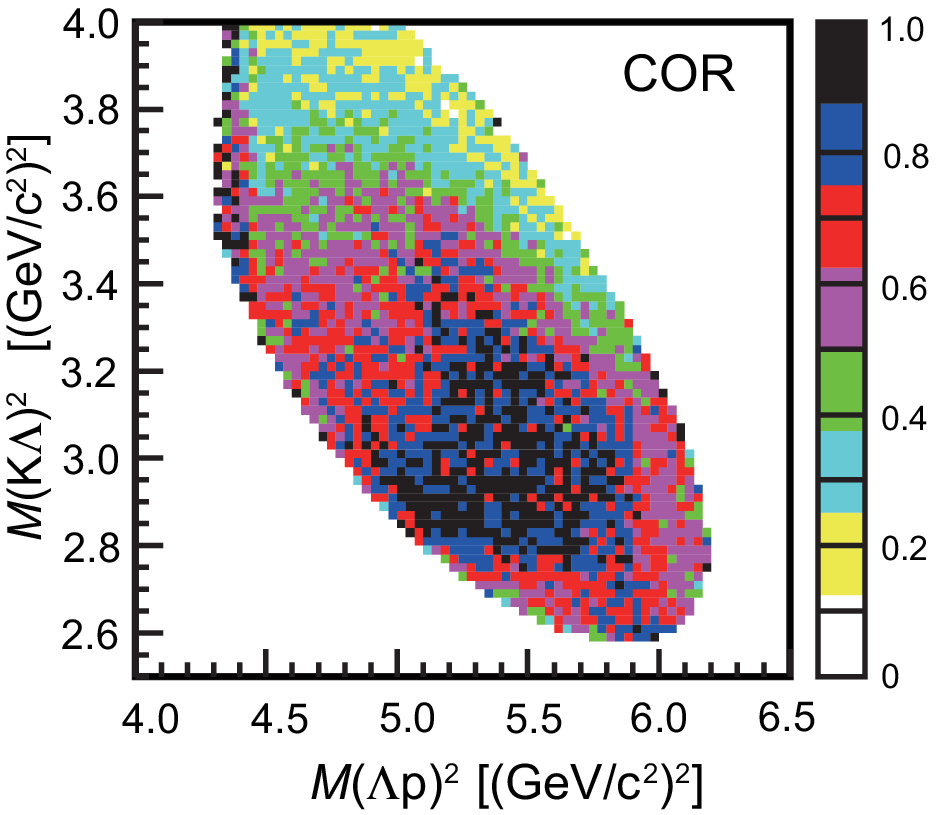}%_DISTO-Dalitz-eff-corr-h
       \caption{\label{fig:DISTO-Dalitz}
(Left) A $\Delta M (p K^+)$ spectrum of raw data after the kinematically constrained refit,
acceptance uncorrected; the small shaded histogram includes events after $p$ and $K^+$ cuts applied to obtain final spectra. (Right) An acceptance-corrected Dalitz plot of the exclusive $pp \rightarrow p \Lambda K^+$  reaction products at 2.85 GeV.     }     
\vspace{-0.5cm}
  \end{center}
\end{figure}

The DISTO experiment, originally aimed at comprehensive studies of strangeness exchange reactions in $pp$ collisions, was carried out at the SATURNE accelerator at Saclay \cite{DISTO-NIM,DISTO}. We have analyzed the experimental data set of exclusive reaction products, $p \Lambda K^+$, at the incident kinetic energy of 2.85 GeV (see details in \cite{Maggiora:09}). For this purpose $\Lambda$ events were preselected by identification of the $\Lambda \rightarrow p \pi^-$ decay, and a missing-mass spectrum of $\Delta M (p K^+)$ showed peaks of $\Lambda$, $\Sigma^0$ and $\Sigma^0(1385) + \Lambda(1405)$. Then, they were used for a kinematical refit of four particles, $p, p, K^+, \pi^-$, where the reconstructed invariant mass $M_{p\pi^-}$ is constrained to the $\Lambda$ mass and the four-body missing mass is constrained to 0 or $M_{\pi^0}$. The reaction products are selected by asking for the Dalitz regions kinematically accessible by the $p \Lambda K^+$ and $p \Sigma^0 K^+$ final states. The missing-mass spectrum, $\Delta M (p K^+)$, thus obtained, is shown in Fig.~\ref{fig:DISTO-Dalitz} (left). We selected about 177,000 exclusive $p \Lambda K^+$ events by setting the $\Lambda$ gate as shown. The impurity of $\Sigma^0$ in this $\Lambda$ gate is estimated to be 4.5 \%. The $\Sigma^0$ impurity is found to be reduced to 2.6 \% for events in final spectra to be shown later.

The exclusive $p \Lambda K^+$ data not only correspond to the ``ordinary" $\Lambda$ production process: 
\begin{equation}
{\rm [ordinary]}~~ p + p \rightarrow p + \Lambda + K^+, \label{eq:pp2LpK}
\end{equation}
but also may include the ``exotic" process involving the $K^-pp$  bound state ($\equiv X$):  
\begin{equation}\label{eq:pp2KX}
{\rm [exotic]}~~ p + p \rightarrow K^+  + X,~~~
         X \rightarrow p + \Lambda.
\end{equation} 
 
\begin{figure}[htb]
\centering
\includegraphics[width=7.5cm]{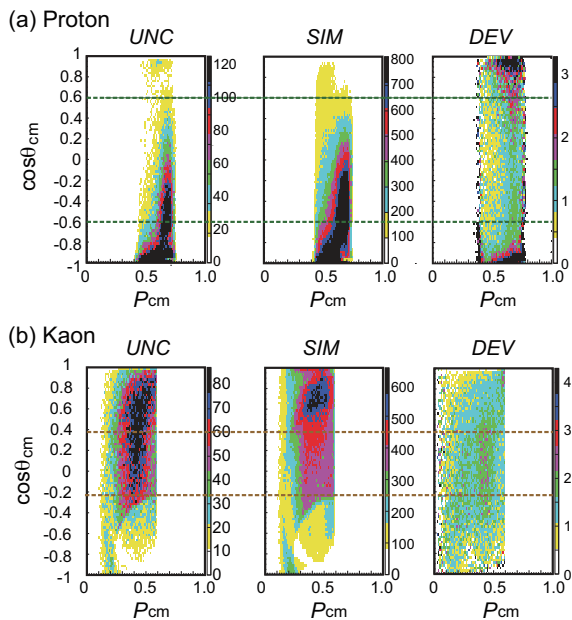}%_pLK-Distribution-e
\vspace{0cm}
\caption{\label{fig:p-K-distribution} 
Momentum distributions, $P$, versus cos$ \, \theta$ in c.m. of (a) $p$ and (b) $K^+$. Each block consists of  {\it UNC} (uncorrected), {\it SIM} (simulated), and {\it DEV} (deviation) data. The horizontal dotted lines define $-0.6 < {\rm cos}\, \theta_{\rm cm} (p) < 0.6$ and $-0.2 < {\rm cos} \, \theta_{\rm cm}(K) < 0.4$, to be used for proton-angle and $K^+$-angle cuts. }
\end{figure}

 We show in Fig.~\ref{fig:DISTO-Dalitz} (right) an acceptance-corrected Dalitz plot of all the $p \Lambda K^+$ events in the plane of $x = M(p \Lambda)^2$ vs $y = M(\Lambda K)^2$. 
 The expected Dalitz distribution of the ``ordinary" process (\ref{eq:pp2LpK}) {\it is continuous without any local bump structure} \cite{Akaishi:08b}. On the other hand, the observed distribution, Fig.~\ref{fig:DISTO-Dalitz} (right), reveals some structure that cannot be explained by the ``ordinary" process. However, the Dalitz plot alone cannot discriminate between the ``ordinary" and the ``exotic" processes. It is important to study the angular distributions and momentum spectra of $p$, $\Lambda$ and $K^+$, with respect to the incident beam, which are hidden in the Dalitz presentation. 

Hereafter, we show {\it uncorrected experimental} spectral distribution of a kind $\alpha$ ({\it UNC\/}$^{(\alpha)}$) together with {\it simulated} data ({\it SIM\/}$^{(\alpha)}$). Each {\it SIM\/}$^{(\alpha)}$ distribution is calculated for events of the ``ordinary" $pK^+\Lambda$ process (\ref{eq:pp2LpK}) according to a uniform phase-space distribution, folded with the DISTO acceptance, and then fed to the complete reconstruction and analysis chain for {\it UNC\/}$^{(\alpha)}$, fulfilling hence the same cuts and refitting procedure. In view of possible uncertainties in the efficiency matrix we adopt a method to obtain an {\it efficiency-compensated} presentation of the experimental data, that is, by calculating a {\it deviation spectrum} for each bin as defined by
\begin{equation}
{\it DEV}^{(\alpha)} \equiv {\it UNC}^{(\alpha)}/{\it SIM}^{(\alpha)}.
\end{equation}
Each {\it DEV\/}$^{(\alpha)}$ spectrum is, of course, different from the intensity distribution, which is generally bell-shaped due to the phase-space density. {\it DEV\/}$^{(\alpha)}$ is free, not only from the phase-space density, but also from possible uncertainty in the efficiency. A {\it DEV\/}$^{(\alpha)}$ distribution is generally flat, but indicates a structure when a physically meaningful deviation from the uniform phase-space distribution occurs. This {\it deviation-spectrum method}  is valid, when the selected events are 100\% exclusive $p \Lambda K^+$ \cite{Maggiora:09}. The actual purity of the $\Lambda$ selection is about 95 \%, and is even higher for events taken in the final spectra. The validity can be examined from the equivalence of $M(p\Lambda)$ and $\Delta M(K^+)$, which are in fact found to be nearly identical to each other. 

Figure \ref{fig:p-K-distribution} shows the {\it UNC}, {\it SIM} and {\it DEV} distributions of the momentum $P$ vs cos$\,\theta$ in c.m. of $p$ and $K^+$. 
The {\it UNC} data as well as the {\it SIM} data show that the $p$ distribution is extremely backward, which arises from the large acceptance of the DISTO detector for $\Lambda \rightarrow p + \pi^-$ decay in the forward direction. The {\it DEV} distributions of $p$ and $\Lambda$ (not shown), both peaked at cos$\,\theta = \pm 1$ in c.m., are remarkably symmetric, as expected from the symmetric $pp$ collision in c.m. This fact justifies the present acceptance correction on {\it SIM}. 

Since the maximum momentum of $p$ in the ``ordinary" process of $p \Lambda K^+$ is 0.751 GeV/$c$, the dominating proton group at extremely backward angles (cos$ \, \theta_{\rm cm} (p) < -0.9$) has a transverse momentum, $P_T < 0.3$ GeV/$c$. This fact is understood as the ``ordinary" process of low transverse momentum $P_T$ \cite{Akaishi:08b}. On the other hand, since the proton group of large-angle emission includes a much less  amount of the ``ordinary" process, the relative contribution of the ``exotic" process, involving the decay of $X$ with a transverse momentum of around 0.4 GeV/$c$, is expected to be larger.  So, it is extremely interesting to distinguish $X$ from the dominant ``ordinary" process by applying proton-angle cuts, such as ``large-angle proton" (LAP) cut: $|$cos$ \, \theta_{\rm cm} (p)| < 0.6$ and ``small-angle proton" (SAP) cut: $|$cos$ \, \theta_{\rm cm} (p)| > 0.6$.

\begin{figure}[htbp]
\centering
\includegraphics[width=7.3cm]{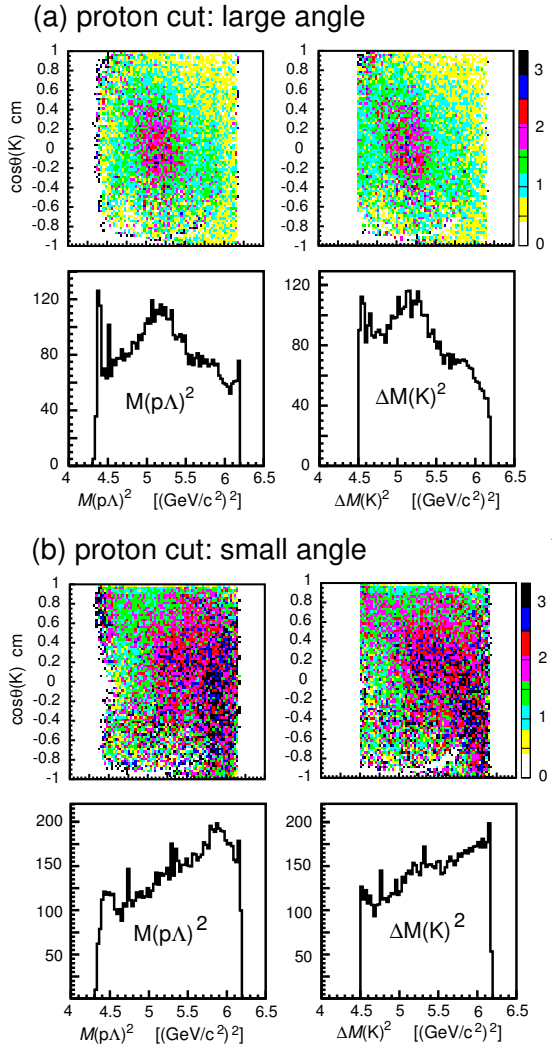}%_Dalitz+K-pcut-b
\vspace{0cm}
\caption{\label{fig:Dalitz+K-pcut} 
(a) Events with LAP cut ($|$cos$\, \theta_{\rm cm} (p) | < 0.6$) and (b) events with SAP cut ($|{\rm cos}\, \theta _{\rm cm} (p)| > 0.6$). Each frame consists of (upper) {\it DEV} spectra of 2-dimensional $M (p \Lambda)^2$ and $\Delta M (K)^2$ vs cos$\, \theta _{\rm cm} (K)$ and (lower) {\it DEV} spectra of 1-dimensional $M (p \Lambda)^2$ and $\Delta M (K)^2$.}
\end{figure}

\begin{figure}[htbp]
\centering
\includegraphics[width=6.8cm]{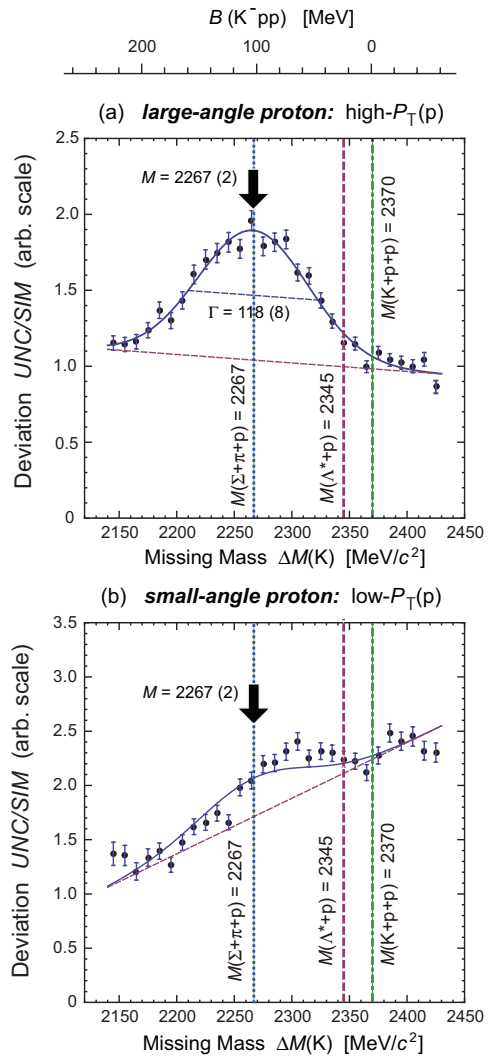}%_MM_p36_notp36-r
\vspace{-0.3cm}
\caption{\label{fig:MM-K} 
(a) Observed {\it DEV} spectra of $\Delta M(K^+)$ of events with LAP emission ($|$cos$\, \theta_{\rm cm} (p) | < 0.6$) and (b) with SAP emission ($|$cos$\, \theta_{\rm cm} (p) | > 0.6$). Both selected with large-angle $K^+$ emission
 ($-0.2 <  {\rm cos}\, \theta_{\rm cm} (K^+)  < 0.4$).} 
\vspace{-0.5cm}
\end{figure}

The {\it DEV} $K^+$ spectrum of Fig.~\ref{fig:p-K-distribution} (b) already shows that the $P_{\rm cm} (K)$ has a monoenergetic component around 0.4 GeV/c even before applying the proton-angle cut, which could be a signature for a two-body process, $pp \rightarrow K^+ X$. It is to be noted that this component is present in the {\it UNC} data, and thus, the component appearing in the {\it DEV} data {\it cannot be a fake} that might originate from the correction, since {\it SIM} is smooth in this region. 

Figure~\ref{fig:Dalitz+K-pcut} shows {\it DEV} spectra of $M (p \Lambda)^2$ and $\Delta M (K)^2$ and their correlations with cos$\, \theta _{\rm cm} (K)$ for (a) LAP and (b) SAP. Clearly, the vertical band in the $P_{\rm cm}(K)$ vs cos$\, \theta_{\rm cm} (K)$ plot, now converted into $M(p\Lambda)^2$ and $\Delta M (K)^2$, is enhanced with LAP. A distinct peak is seen at around $x \sim 5.15$, corresponding to $M_X \approx 2.27$ GeV/$c^2$. We have demonstrated that the proton-angle cut is very effective in discriminating the $X$ formation process (monoenergetic emission of $K^+$) from the ``ordinary" process. The {\it DEV} spectra of SAP, as shown in Fig.~\ref{fig:Dalitz+K-pcut} (b), are remarkably flat. The linear shape with a positive gradient of (b) is well accounted for by the ``ordinary" $pp \rightarrow p \Lambda K^+$ mechanism with a collision length of $\hbar/m_B c$ with $m_B \sim 0.2$ GeV/$c^2$, which explains the proton angular distribution as well \cite{Akaishi:08b}. In great contrast, the observed {\it DEV} spectra of LAP (a) reveal a large bump on a flat horizontal background.

We find from Fig.~\ref{fig:p-K-distribution} (b) that the geometrical acceptance of $K^+$ is not flat in the forward and backward c.m. zones of $K^+$ emission, as seen in both {\it UNC} and {\it SIM}. Thus, we applied cuts, $-0.2 < {\rm cos} \, \theta_{\rm cm} (K) < 0.4$, to obtain final {\it DEV} spectra, where enhanced monoenergetic $X$-events are observed at large $K^+$ c.m. angles.

In summary, 
 we have observed a large broad peak in the {\it DEV} spectrum of $\Delta M (K^+)$ (and $M(p \Lambda)$), as shown in Fig.~\ref{fig:MM-K} (a), which is associated with LAP. On the other hand, in Fig.~\ref{fig:MM-K} (b) for SAP,
 a linear background dominates. 
 We made simple fitting of the {\it DEV} spectrum of $\Delta M (K^+)$ with LAP (a) by a Gaussian + linear background, and obtained 
\begin{eqnarray}
 M_X &=& 2265 \pm 2~(stat) \pm 5~(syst)  ~{\rm MeV}/c^2,\\
 \Gamma_X  &=& 118 \pm 8~(stat) \pm 10~(syst) ~{\rm MeV}.
\end{eqnarray}
The best-fit $\chi ^2 /ndf$ value is 34.2/24 = 1.4. The peak height amounts to about 26-$\sigma$ statistical significance. The {\it DEV} spectrum of $\Delta M (K^+)$ with SAP (b) can be fitted with the same Gaussian shape and another linear background, as shown. The very different background slopes of (a) and (b) are accounted for by a simple reaction model \cite{Akaishi:08b}. The Gaussian-like peak persists in (b) despite the larger background for SAP. The observed mass of $X$ is close to $M(p \Lambda) \sim$ 2255 MeV/$c^2$ of the $K^-pp$ candidate reported by FINUDA \cite{Agnello:05}. 
 
The $X$ production rate is found to be as much as the $\Lambda(1405) (= \Lambda^*)$ production rate, which is roughly 20 \% of the total $\Lambda$ production rate. Such a large formation is theoretically possible {\it only when the $p$-$p$} (or $\Lambda^*$-$p$) {\it rms distance in $X$ is shorter than 1.7 fm} \cite{Yamazaki:07a,Yamazaki:07b}, whereas the average $N$-$N$ distance in ordinary nuclei is 2.2 fm. The $pp \rightarrow \Lambda^* + p + K^+ \rightarrow X + K^+$ reaction produces $\Lambda^*$ and $p$ of large momenta, which can match the internal momenta of the off-shell $\Lambda^*$ and $p$ particles in the bound state of $X = \Lambda^*$-$p$, only if $X$ exists as a dense object. Thus, the dominance of the formation of the observed $X$ at high momentum transfer ($\sim 1.6$ GeV/$c$) gives direct evidence for its compactness of the produced $K^-pp$ cluster.

The observed mass of $X$ corresponds to a binding energy $B_K = 105 \pm 2~(stat) \pm 5~(syst)$ MeV for $X = K^-pp$. It is larger than the original prediction \cite{Yamazaki:02,Shevchenko,Ikeda}. It could be accounted for, if the $\bar{K}N$ interaction is effectively enhanced by 25 \%, thus suggesting additional effects to be investigated \cite{Yamazaki:07b,Wycech}. On the other hand, the theoretical claims for shallow $\bar{K}$ binding \cite{Oset,Weise:07,Dote:09} do not seem to be in agreement with the present observation.

As shown in Fig.~\ref{fig:MM-K}, the peak is located nearly at the $\Sigma \pi$ emission threshold, below which the $N \Sigma \pi$ decay is not allowed. The expected partial width, $\Gamma_{N \Sigma  \pi}$, must be much smaller than the predicted value of 60 MeV \cite{Yamazaki:02}, when we take into account the pionic emission threshold realistically by a Kapur-Peierls procedure (see \cite{Akaishi:08}). Thus, $\Gamma_{\rm non-\pi} = \Gamma_{p\Lambda} + \Gamma_{N \Sigma} =  \Gamma_{\rm obs} - \Gamma_{N \Sigma  \pi} \approx 100$ MeV, which is much larger than recently calculated non-pionic widths for the normal nuclear density, $\Gamma_{\rm non-\pi} \sim 20 - 30$ MeV \cite{Sekihara,Ivanov:09,Wycech}. The observed enhancement of $\Gamma_{\rm non-\pi}$ roughly by a factor of 3 seems to be understood with the compact nature of $K^-pp$  \cite{Yamazaki:07b}. 

We are indebted to the stimulating discussion of Professor Y. Akaishi and R.S. Hayano. This research was partly supported by the DFG cluster of excellence ``Origin and Structure of the Universe" of Technische Universit\"at M\"unchen and by Grant-in-Aid for Scientific Research of Monbu-Kagakusho of Japan. One of us (T.Y.) acknowledges the support by an Award of the Alexander von Humboldt Foundation, Germany.

\end{document}